\documentclass[final,5p,times,twocolumn,nopreprintline]{elsarticle}
\usepackage{amsmath,amssymb,amsfonts}
\usepackage{graphicx,subcaption,epsfig}
\usepackage{xcolor}
\usepackage{dcolumn}
\usepackage{multicol}
\usepackage{upquote,multirow,slashed}
\usepackage{appendix}
\usepackage{empheq}
\usepackage{lipsum}
\usepackage{booktabs,dsfont}
\usepackage{dcolumn}
\usepackage{xspace}
\usepackage{balance}

\definecolor{ForestGreen}{rgb}{0.15,0.70,0.15}
\usepackage[colorlinks=true]{hyperref}
\hypersetup{
 linktocpage=true,
 citecolor=ForestGreen,
 linkcolor=blue,
 urlcolor=blue}

\usepackage{fancyhdr}
\fancypagestyle{firstpage}{%
	
	\lhead{}
	\rhead{\small CPTNP-2025-015, MPP-2025-111}
}
\allowdisplaybreaks[1]

\newcommand{\NOdisplay}[1]{ }

\def\MSbar{\overline{\mathrm{MS}}}

\begin{document}

% ******************************* 
% Title & details of the authors
% *******************************

\begin{frontmatter}

\title{
Three-loop QCD Mass Relation between the $\overline{\mathrm{MS}}$ and Symmetric-momentum Subtraction Scheme Away from the Chiral Limit
}

\author[SDU]{Long Chen}
\ead{longchen@sdu.edu.cn}
\author[MPI]{Marco Niggetiedt}
\ead{marco.niggetiedt@mpp.mpg.de}

\address[SDU]{School of Physics, Shandong University, Jinan, Shandong 250100, China}
\address[MPI]{Max-Planck-Institut f\"ur Physik, Boltzmannstra{\ss}e~8, 85748 Garching, Germany}

\begin{abstract}
The perturbative result for the quark-mass conversion factor between the $\overline{\mathrm{MS}}$ and regularization-independent symmetric-momentum subtraction scheme (RI/SMOM) away from the chiral limit, i.e.~at non-zero quark masses (RI/mSMOM), is derived up to three loops in QCD, extending the existing result by two additional orders. 
We further explore an illuminating possibility that in Dimensional Regularization, the original RI/(m)SMOM renormalization conditions may be interpreted merely in a weaker sense, namely as equations holding just in the 4-dimensional limit rather than exactly in $d$ dimensions: they result in different, albeit simpler, renormalization constants but still the same finite conversion factor. This novel observation has the added benefit of reducing computational effort, particularly at high orders.
Our high-order results for the conversion factor exhibit rich behaviors, and in particular a window is observed in the subtraction scale and mass where it receives less perturbative corrections than the RI/SMOM counterpart up to three loops; this finding may help to further improve the accuracy of $\overline{\mathrm{MS}}$ quark-mass determinations with Lattice QCD.
~\\

\end{abstract}

\end{frontmatter}

\thispagestyle{firstpage}

\section{Introduction}
\label{sec:intro}

Quark masses are fundamental parameters of the Standard Model (SM) of particle physics, and their precise values are of utter importance for various high-precision tests of the SM, and accordingly for placing constraints on New Physics.
Lattice QCD (LQCD) has proven to be an increasingly important approach in quark-mass determinations, especially for light quarks with $m_q < \Lambda_{QCD}$ (see, e.g., the recent FLAG Review~\cite{FlavourLatticeAveragingGroupFLAG:2024oxs}), as it allows \textit{ab initio} calculations of the non-perturbative hadronic dynamics involved.
LQCD determinations of heavier quark masses $m_Q \gg \Lambda_{QCD}$ are also highly desirable, and have been successfully done for the $c$-quark mass by several Lattice groups~\cite{HPQCD:2008kxl,Davies:2009ih,McNeile:2010ji,FermilabLattice:2014tsy,Yang:2014sea,Lytle:2018evc,CLQCD:2024yyn,DelDebbio:2024hca}.
The matching relations connecting the lattice-spacing $a$ dependent bare-quark masses $m_{B}(a)$ to a phenomenologically useful (renormalized) mass defined in continuum QCD, in particular the commonly used $\MSbar$-mass $\overline{m}_R$ can be derived, e.g., using Lattice perturbation theory~\cite{Capitani:2002mp,HPQCD:2004hdp,Mason:2005bj}, or regularization-independent (RI) renormalizations~\cite{Martinelli:1994ty,RBC-UKQCD:2008mhs,Aoki:2007xm,Sturm:2009kb,Boyle:2016wis} as well as the current-current correlator method~\cite{HPQCD:2008kxl,McNeile:2010ji}. 
Many of the latest LQCD determinations of quark masses~\cite{RBC-UKQCD:2008mhs,Yang:2014sea,Bi:2017ybi,Lytle:2018evc,Hatton:2020qhk,Hatton:2021syc,He:2022lse,CLQCD:2024yyn,DelDebbio:2024hca} employ certain RI renormalization schemes.
This was suggested initially in ref.~\cite{Martinelli:1994ty}, proposing a RI momentum-subtraction scheme (RI/MOM) that can be implemented both in LQCD and in continuum QCD with Dimensional Regularization (DR), in order to circumvent the computational difficulties with Lattice perturbation theory~\cite{Mason:2005bj,Mason:2005de}. 
To mitigate the technical issues in the practical implementation of RI/MOM related to its choice of an \textit{exceptional} momentum configuration in the renormalization of quark bilinear operators (with zero momentum insertion),
refs.~\cite{Aoki:2007xm,Sturm:2009kb} proposed a modification by allowing a non-zero momentum insertion, 
with virtuality equal to those of the external quark momenta (and still with massless quark propagators), 
which is known as the symmetric momentum-subtraction scheme (RI/SMOM or SMOM).
On the other hand, the perturbative calculation of the mass conversion factor becomes more complicated compared to the RI/MOM case~\cite{Franco:1998bm,Chetyrkin:1999pq,Gracey:2003yr,Chetyrkin:2000fd,Chetyrkin:2008jk,Skouroupathis:2007jd,Skouroupathis:2008mf,Constantinou:2016ieh}, 
and the results up to three loops became available just a few years ago~\cite{Sturm:2009kb,Gorbahn:2010bf,Almeida:2010ns,Kniehl:2020sgo,Bednyakov:2020ugu}.

One important aspect in the LQCD-based determination of heavy quark masses ($m_Q \gg \Lambda_{\mathrm{QCD}}$) is to have a good control over the discretization errors in the extrapolation to the continuum limit ($a \rightarrow 0$), especially those caused by heavy-quark masses~\cite{Lytle:2018evc,He:2022lse,CLQCD:2024yyn,DelDebbio:2024hca}.
In this respect, a natural extension of SMOM was proposed for bilinear quark operators in ref.~\cite{Boyle:2016wis}, termed as RI/mSMOM or mSMOM for short, which retains the benefits of the symmetric momentum configuration and preserves the vector-current and non-anomalous axial-current Ward-Takahashi identities~\cite{Adler:1969gk} exactly at any finite quark masses, i.e. away from the chiral limit.
This scheme can thus be expected to reabsorb some $\mathcal{O}(a^2 m^2_Q)$ discretization artifacts~\cite{Boyle:2016wis}, hence leading to a milder and more reliable continuum extrapolation as compared to SMOM, which was recently confirmed in its first LQCD application in ref.~\cite{DelDebbio:2024hca}. 
However, due to keeping quark propagators massive, the corresponding mass conversion factor 
becomes significantly more involved compared to the SMOM case, hence, its perturbative expression is so-far only known to one loop in ref.~\cite{Boyle:2016wis}.
In the determination of the $c$-quark $\MSbar$ mass in ref.~\cite{DelDebbio:2024hca}, an estimate of the systematic error from this one-loop truncation was based on the difference between the one- and two-loop SMOM results~\cite{Sturm:2009kb,Gorbahn:2010bf,Almeida:2010ns}, to be about $0.4\%$ at the commonly chosen matching scale $2\,$GeV for this problem,\footnote{We note that the three-loop correction to the SMOM mass conversion factor determined in refs.~\cite{Kniehl:2020sgo,Bednyakov:2020ugu} is, in fact, similar in size to the two-loop correction at $2\,$GeV.} which comprises a considerable portion of the total error budget.
Furthermore, the current leading systematic uncertainty in ref.~\cite{DelDebbio:2024hca} is expected to be significantly reduced by using additional finer lattice ensembles, rendering the computation of the mSMOM mass conversion factor to higher loop orders even more desirable.

In this work we close this gap by providing the first determination of the quark-mass conversion factor between the $\MSbar$ and mSMOM up to three loops, overcoming the formidable challenge of high-precision computations of three-loop massive off-shell vertex functions. 
Moreover, several novel and compelling observations are made along the way, which may not only help to refine the application of the mSMOM, but shall also be useful in reducing considerably the computational effort, especially when incorporating additional quarks with different masses.

\section{RI/mSMOM conditions for quark mass}
\label{sec:RCmass}

The relation between the $\MSbar$-renormalized quark mass $\overline{m}_R$ and the corresponding mass $m_R$ defined in mSMOM scheme can be established in dimensionally-regularized continuum QCD via the following multiplicative relation with the bare mass:% 
\begin{equation} \label{eq:massrelation2bare}
m_B = Z_{\overline{m}}(\bar{\mu}) \, \overline{m}_R(\bar{\mu})  
= Z_m(\mu_s) \, m_R(\mu_s)
\end{equation}%
where $\bar{\mu}$ denotes the scale of both $\MSbar$-renormalized $\overline{m}_R(\bar{\mu})$ and the perturbative $\alpha_s(\bar{\mu})$\footnote{The definition of $\bar{\mu}$ follows the usual $\MSbar$ renormalization of the bare coupling: $\alpha^0_s = \alpha_s(\bar{\mu})\, Z_{a_s} \, \bar{\mu}^{2\epsilon} (4 \pi)^{-\epsilon} e^{\epsilon \gamma_E} $ with the DR regulator $\epsilon = (4-d)/2$.}, while $\mu_s$ denotes the momentum subtraction scale in the mSMOM scheme.
Here and below we use symbols with subscript $B(R)$ to denote bare (renormalized) quantities, and for generic objects if this subscript is suppressed.
In perturbative continuum QCD, both $Z_{\overline{m}}(\bar{\mu})$ and $Z_m(\mu_s)$ are determined as functions of $\MSbar$-renormalized $\alpha_s(\bar{\mu})$; 
$Z_m(\mu_s)$ also depends on the non-zero quark-propagator masses involved in the problem, while $Z_{\overline{m}}(\bar{\mu})$ is independent of any masses.
Consequently, one defines the mass conversion factor:%
\begin{equation} \label{eq:massconversionrelation}
\overline{m}_R(\bar{\mu}) = \frac{Z_m(\mu_s) }{Z_{\overline{m}}(\bar{\mu})} \, m_R(\mu_s)
\equiv \mathrm{C}_m(\bar{\mu}, \mu_s) \, m_R(\mu_s)
\end{equation}%
which we determine for the first time up to three-loop order, %i.e.~to $\mathcal{O}(\alpha_s^3)$, 
in continuum QCD with one massive quark, with renormalized mass $m_R$, and $n_l$ massless quarks.
(In the following text we will omit, by default, the arguments of these notations within brackets for simplicity, unless they are needed to ensure definiteness or prevent ambiguity.)

The full set of RI/mSMOM renormalization conditions were specified originally in ref.~\cite{Boyle:2016wis} that involve the 2-point quark propagating function $S(p, m)$ at spacelike kinematics $p^2=-\mu_s^2$ 
and 3-point one-particle-irreducible (1PI) vertex functions for elementary bilinear quark operators $\hat{\mathcal{O}}_{\Gamma}$ (such as vector- and axial-current, scalar and pseudoscalar operators) at the so-called symmetric-momentum configuration. %, which we refrain from repeating here. 
We note that $Z_m$, together with the related quark-field mSMOM renormalization constant $Z_q(\mu_s)$ defined by $\psi_B = Z_q^{1/2}(\mu_s) \, \psi_R$, %\footnote{In our convention, the quark mass and field renormalization constants are the inverse of the respective ones defined in ref.~\cite{Boyle:2016wis}.}, 
can be determined by considering the reduced set of two coupled renormalization conditions:
\begin{eqnarray} 
\frac{1}{4 N_c \, p^2} \mathrm{Tr}\big[i\, S^{-1}_R(p, m_R)\, \slashed{p}\big]\Big|_{p^2 = -\mu_s^2}\, =\, 1\,, \label{eq:massRCs_f}
\\
\frac{1}{4 i N_c} \mathrm{Tr}\big[ \Lambda_{P\,,R}\, \gamma_5\big]\Big|_{p_1^2 = p_2^2 = q^2 = -\mu_s^2}\, =\, 1\,, \label{eq:massRCs_m}
\end{eqnarray}
with the renormalized quark-propagator mass set at $m_R$, %for the massive quark propagators 
provided the validity of the vector-current and non-anomalous (non-singlet) axial-current Ward-Takahashi identities~\cite{Adler:1969gk} in QCD at the bare level in DR (which ensures the equality $Z_P = Z_m$ in mSMOM~\cite{Boyle:2016wis}). %Note that these two conditions take exactly the same form in SMOM and mSMOM.
In eq.~\eqref{eq:massRCs_f}, $S_R(p, m_R)$ denotes the renormalized quark propagating function with off-shell momentum flow $p^2=-\mu_s^2$. 
In eq.~\eqref{eq:massRCs_m}, $\Lambda_{P\,,R}$ denotes the 3-point 1PI vertex function for the renormalized pseudoscalar operator $\hat{\mathcal{O}}_{P\,,R} = Z_{P} \, \hat{\mathcal{O}}_{P\,,B} = Z_{P}\, \bar{\psi}_B i \gamma_5 \psi_B$ with momentum insertion $q = p_1+p_2$ at the symmetric momentum configuration $p_1^2 = p_2^2 = q^2 = -\mu_s^2$.\footnote{Note that here we consider only one massive quark flavor in the definition of this pseudoscalar operator. Summing $\bar{\psi}_B i \gamma_5 \psi_B$ over all quark flavors with equal weight gives the complete flavor-singlet pseudoscalar operator. 
}

Unlike in SMOM, we do not use the simpler scalar operator $\hat{\mathcal{O}}_{S,R}  = Z_S\, \hat{\mathcal{O}}_{S,B} = Z_S\, \bar{\psi}_B\, \psi_B$ for the mass renormalization, as we could not identify a simple all-order renormalization condition for $\hat{\mathcal{O}}_{S,R}$ at $p_1^2 = p_2^2 = q^2 = -\mu_s^2$ with a non-zero propagator mass that could lead to $Z_S = Z_P$. 
(See more comments on this in section~\ref{sec:rest_acrobatic}.)

Due to the off-shell momentum configuration, both $S(p, m)$ and $\Lambda_{P,B}$ become dependent on the gauge-fixing condition. %in the perturbative theory.
There thus rises an additional freedom related to the renormalization of the general covariant-gauge-fixing parameter $\xi$ in the perturbative calculation of the $\xi$-dependent $\mathrm{C}_m$ beyond one loop in QCD, %where this renormalization is needed, 
because the $\xi$-renormalization is not fixed by the mSMOM renormalization conditions specified in ref.~\cite{Boyle:2016wis}.
Fortunately, in the Landau gauge-fixing condition, the multiplicative renormalization of $\xi$ happens to be absent as it corresponds to $\xi=0$.
The LQCD calculation in ref.~\cite{DelDebbio:2024hca} using mSMOM was performed with Landau gauge-fixing. %, just as in the other LQCD calculations done using SMOM scheme.
Hence, for the numerical results to be discussed in section~\ref{sec:rest}, we will only consider Landau gauge ($\xi = 0$). 
On the other hand, we have kept the full $\xi$-dependence in our calculations. 
For $\xi \neq 0$, we take the usual choice of setting the renormalization constant for $\xi$ to be the same as that of gluon field in order to keep the form of gauge-fixing term un-renormalized. %(a typical choice in perturbative calculations).
For the moment, the $\MSbar$ renormalization of the gluon field is employed, similar as in ref.~\cite{Almeida:2010ns}, and the result for $\mathrm{C}_m$ with so-determined $\xi$-dependence is provided in the supplemental file. %associated with this work.
%The result in the Landau gauge, which is eventually needed to convert the mSMOM masses determined in ref.~\cite{DelDebbio:2024hca} to the $\xi$-independent $\MSbar$ mass, can be obtained by setting $\xi = 0$.
We note, however, that the $\MSbar$ renormalization of the gluon field cannot be implemented in LQCD calculation. 
To amend this it shall be feasible to supplement the existing mSMOM conditions~\cite{Boyle:2016wis} with an additional RI renormalization condition for the gluon field in an analogue to eq.~\eqref{eq:massRCs_f}, namely demanding the full renormalized gluon propagator at the spacelike kinematics $p^2 = -\mu_s^2$ to coincide with its tree-level form.
We leave the investigation of such a possibility, and derivation of the corresponding $\xi$-dependent mass conversion factor, for a future work.

\section{Computation details}
\label{sec:tech}

\begin{figure}[htbp]
  \centering
  \begin{subfigure}[b]{0.15\textwidth}
    \centering
    \includegraphics[scale=0.13]{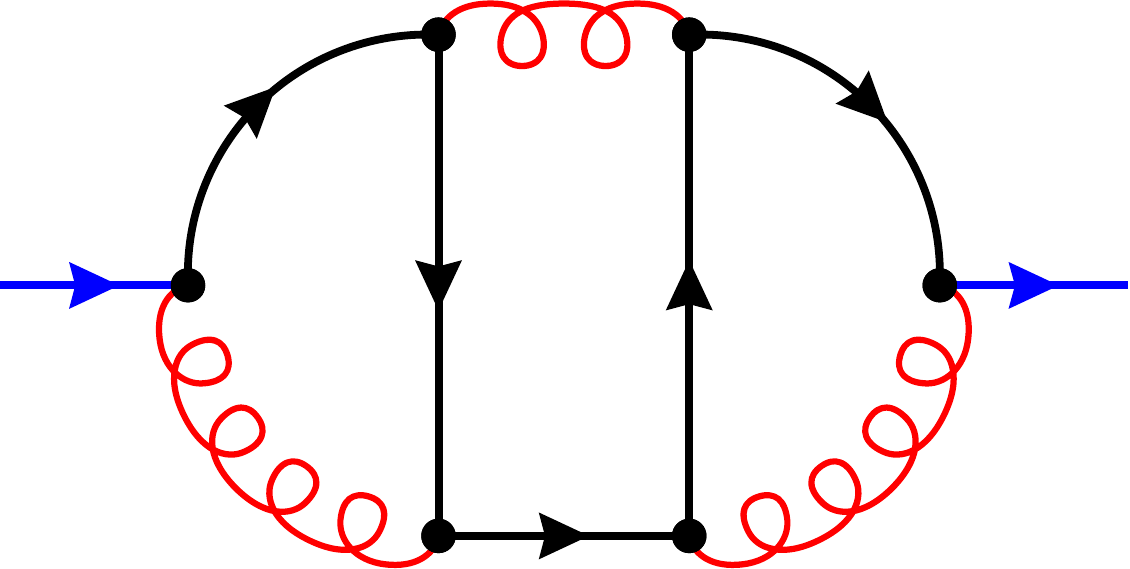}
    \caption{} \label{fig:prop:1}
  \end{subfigure}%
  \begin{subfigure}[b]{0.15\textwidth}
    \centering
    \includegraphics[scale=0.13]{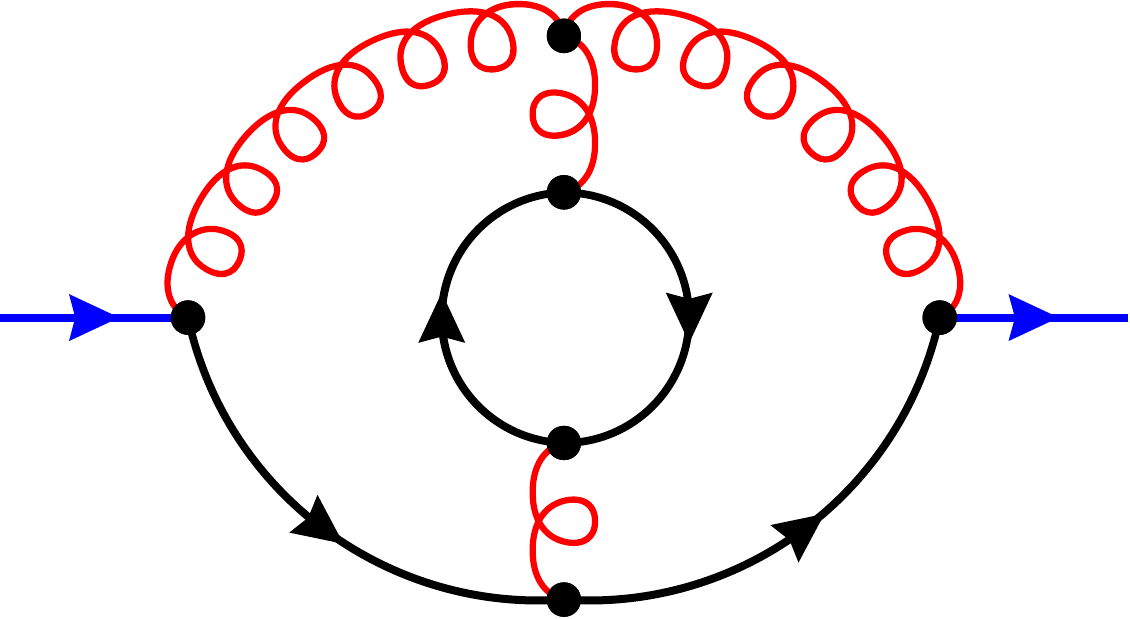}
    \caption{} \label{fig:prop:2}
  \end{subfigure}%  
  \begin{subfigure}[b]{0.15\textwidth}
    \centering
    \includegraphics[scale=0.13]{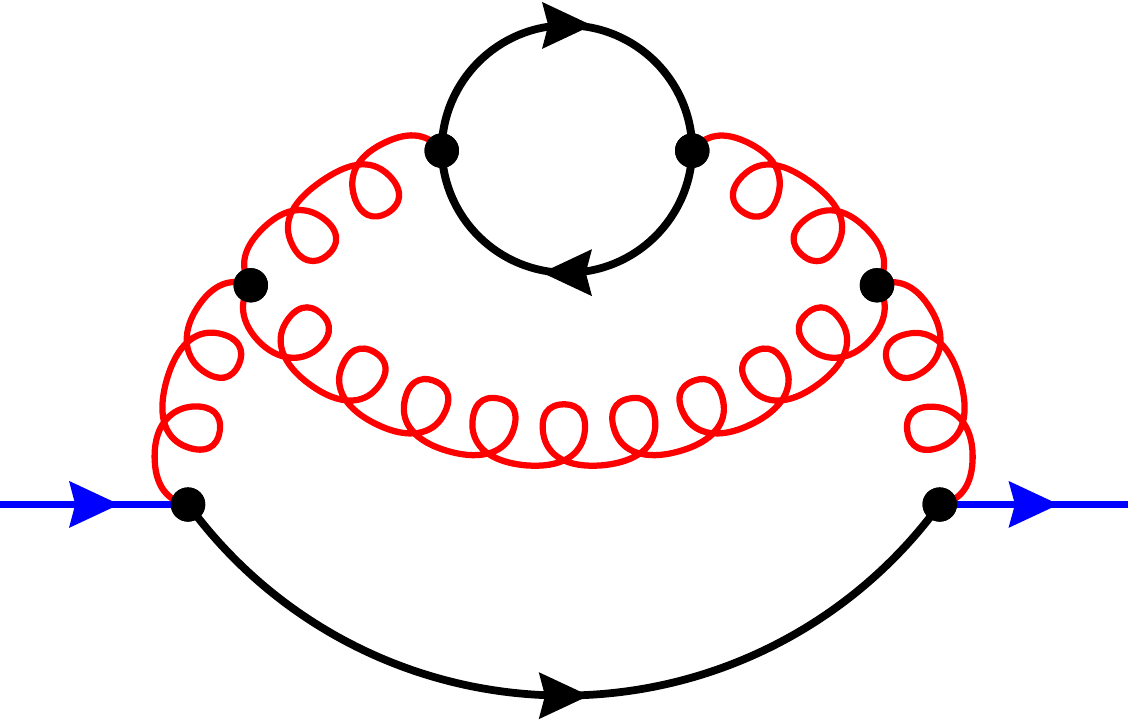}
    \caption{} \label{fig:prop:3}
  \end{subfigure}%  
  \\[14pt]
  \centering
  \begin{subfigure}[b]{0.15\textwidth}
    \centering
    \includegraphics[scale=0.13]{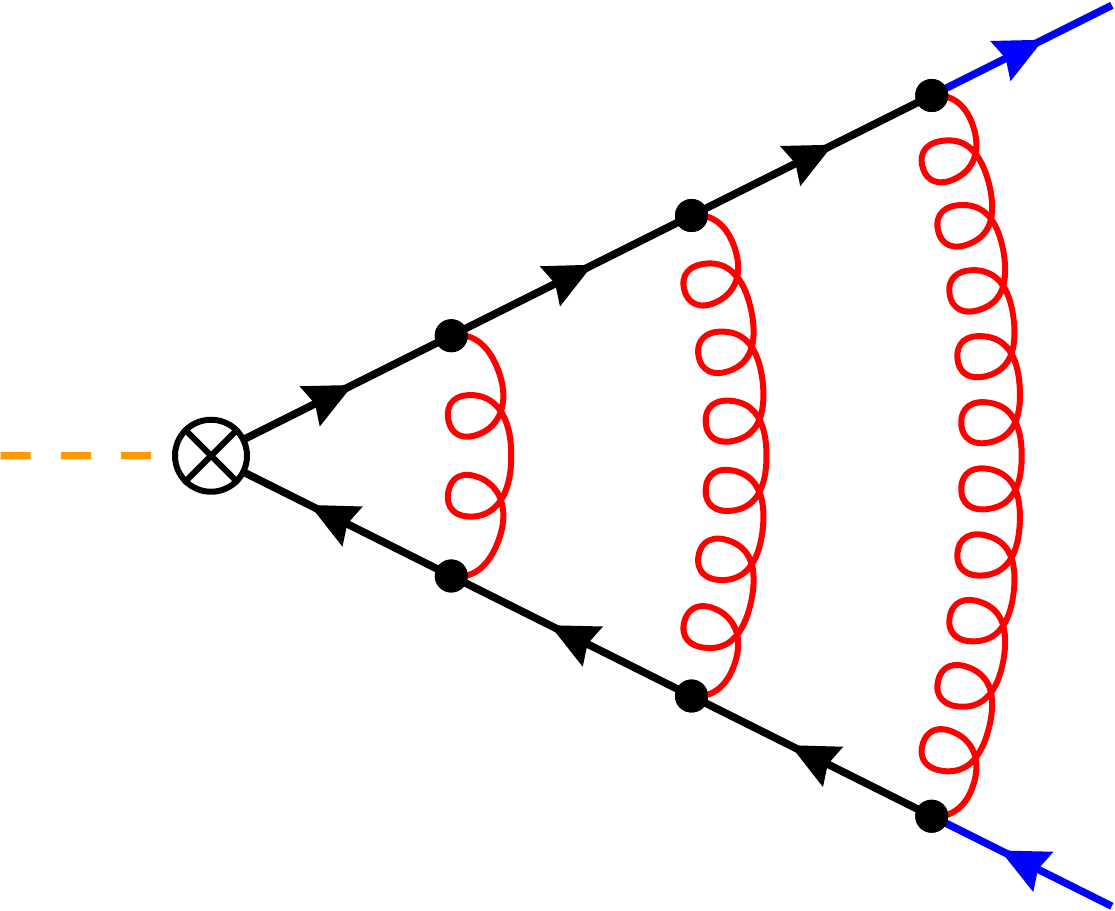}
    \caption{} \label{fig:vert:1}
  \end{subfigure}%
  \begin{subfigure}[b]{0.15\textwidth}
    \centering
    \includegraphics[scale=0.13]{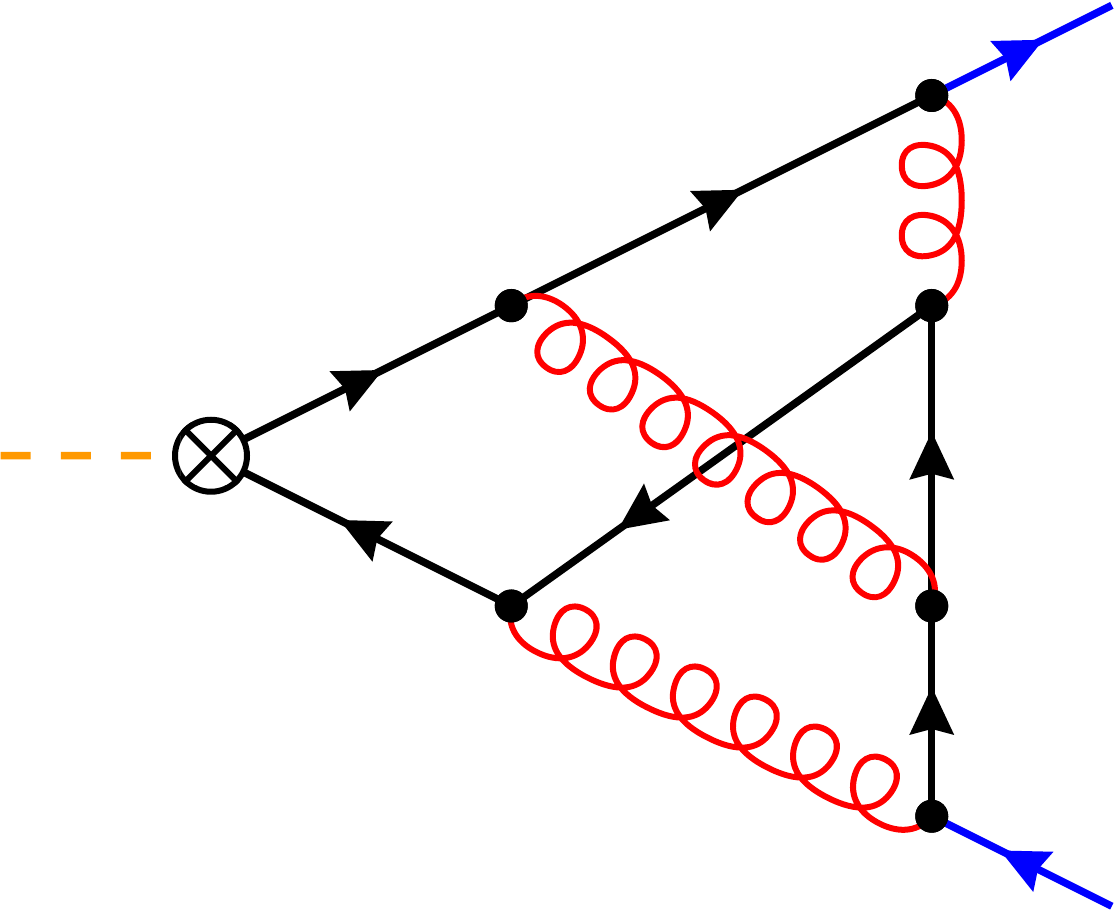}
    \caption{} \label{fig:vert:2}
  \end{subfigure}% 
  \begin{subfigure}[b]{0.15\textwidth}
    \centering
    \includegraphics[scale=0.13]{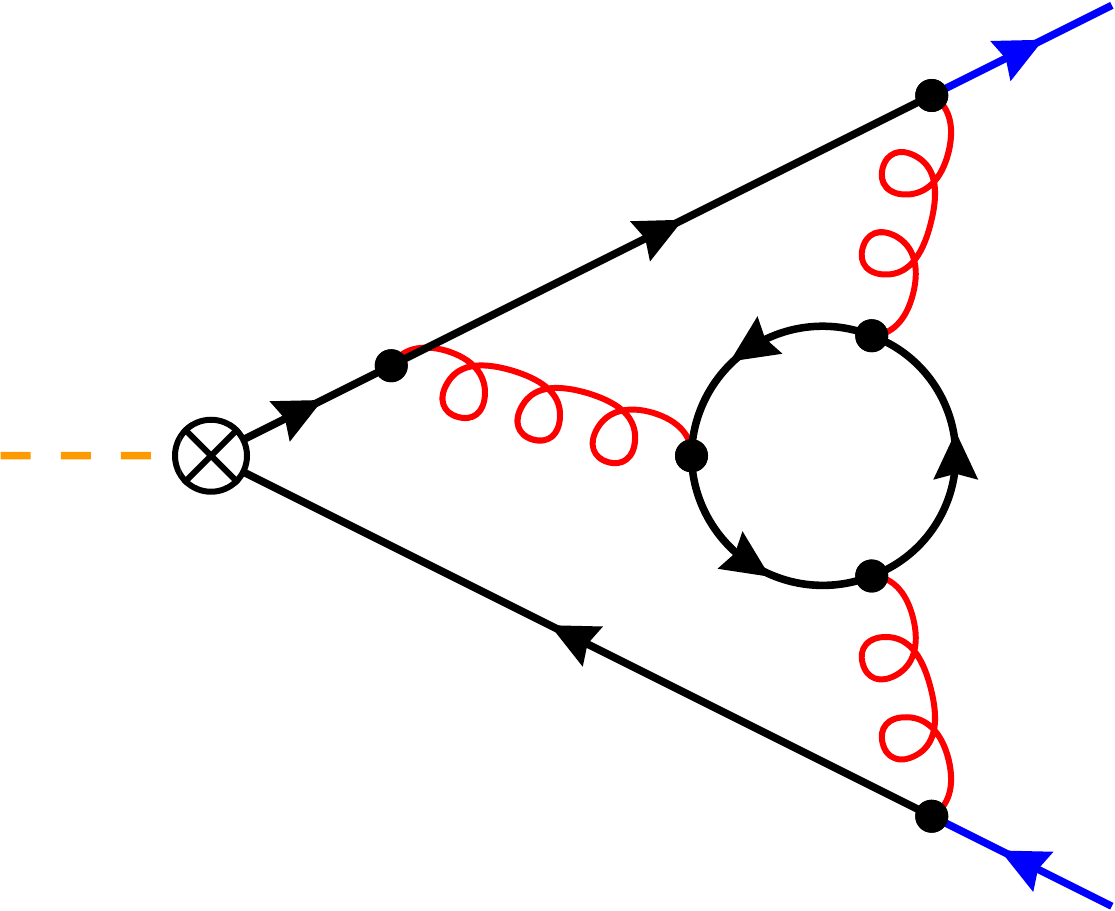}
    \caption{} \label{fig:vert:4}
  \end{subfigure}% 
  \caption{Representative Feynman diagrams at three loops}
    \label{fig:feyndiags}
\end{figure}

We work in continuum perturbative QCD with one massive quark and $n_l$ massless quarks, 
regularized in $d = 4 - 2\epsilon$ dimensions, and compute the perturbative corrections to $\mathrm{C}_m$ defined in~\eqref{eq:massconversionrelation} to $\mathcal{O}(\alpha_s^3)$. 
The perturbative corrections to $S(p, m)$ in~\eqref{eq:massRCs_f} and $\Lambda_{P}$ in~\eqref{eq:massRCs_m} are generated in terms of Feynman diagrams using \texttt{DiaGen}. %~\cite{DiaGen}. 
A few representative three-loop diagrams are illustrated in figure~\ref{fig:feyndiags}. 
In the present work, for $\Lambda_{P}$ we take into account only the contributions of the so-called non-singlet-type diagrams where the external $\mathcal{O}_P$ is attached directly to the open fermion line with a pair of external quarks assigned with equal mass.\footnote{%In other words, this effectively amounts to considering a flavor-changing bilinear quark operator where the two quark fields in the operator have different flavors, hence never become closed under flavor-preserving QCD interactions. %in spite of their "accidental" equal masses (namely this external operator is a flavor-changing/mixed operator).
The so-called singlet-type Feynman diagram contribution to $\Lambda_{P}$, where the pseudoscalar operator couples to a closed fermion chain, starts from two-loop order. However, we have checked that there is no contribution of this type to $\mathrm{C}_m$ at two-loop order. We further observe that there is no contribution of this type to $Z_q$ up to three-loop order.}
Consequently, $\gamma_5$ in $\Lambda_{P}$ of~\eqref{eq:massRCs_m} can be manipulated straightforwardly in a fully anti-commuting manner~\cite{Bardeen:1972vi,Chanowitz:1979zu,Gottlieb:1979ix}.
The Lorentz/Dirac and color algebra are performed with \texttt{FORM}~\cite{Vermaseren:2000nd,Kuipers:2012rf,Ruijl:2017dtg}.%and the package~\texttt{color}~\cite{vanRitbergen:1998pn}.%
%
%Although mSMOM is shown to be beneficial in reducing systematic errors related to certain heavy-quark-mass dependent discretization effects in lattice calculation~\cite{DelDebbio:2024hca}, there is, however, price to pay, from the side of perturbative continuum QCD:
The choice of an off-shell momentum subtraction point with massive quark propagators in the mSMOM conditions~\eqref{eq:massRCs_f} and~\eqref{eq:massRCs_m} leads to loop integrals extremely challenging to compute especially at high loop orders, i.e.~much more complicated than those in RI/MOM and SMOM schemes. %(where the subtraction point is eventually implemented in the chiral limit). 
We note that the diagrammatic Large-Mass-Expansion (LME) method~\cite{Gorishnii:1989dd,Smirnov:1990rz,Smirnov:1994tg} becomes directly applicable to the loop integrals at this configuration. %\footnote{In contrast, this is not feasible for the on-shell massive quarks unless some auxiliary large mass parameters are introduced}.
Consequently, boundary conditions can be obtained algorithmically %(even in exact numbers in most of them in our calculation), 
and a power-logarithm series can thus be constructed with the aid of the differential equations (DEs) in the quark mass, essentially to any finite depth limited only by the capacity of the available computing facilities and time-constraint.
%In particular, for the non-singlet diagrams in question, there is only one non-zero contributing region, the hard region, in LME limit.
This allows us to quickly validate the correct renormalization behavior of our bare expressions for all quark self-energy functions and amputated vertex functions up to three-loop order, as well as the mSMOM renormalization conditions~\cite{Boyle:2016wis} (except for the one for the scalar operator, see section~\ref{sec:rest_acrobatic} for more comments).

We now summarize a few technical details of our computation of the off-shell vertex loop integrals at three loops (and leave a more detailed exposition to a future work taking into account the complete set of singlet diagrams).
After reduction of loop integrals accomplished with \texttt{Kira}~\cite{Maierhofer:2017gsa,Klappert:2020nbg} in combination with \texttt{FireFly}~\cite{Klappert:2019emp,Klappert:2020aqs}, 
we end up with 1115 master integrals, which is more than twice of those encountered in the cutting-edge calculation of on-shell heavy-quark form factors at three-loop order~\cite{Fael:2022rgm,Chen:2022vzo}. %the non-singlet part.
The number of three-loop masters in the massless limit is about 60~\cite{Kniehl:2020sgo}.
The DEs for our optimized basis of masters with symbolic dependence on $\epsilon$ and $m_s \equiv m_R^2/\mu^2_s$ are about 280~MB in size. 
%Although, no full reduction of the differential equations for the naive basis has been carried out, we suspect a significant reduction in complexity as the size of differential equations for the naive basis at fixed ms amounts to approximately 350 MB.
In the top-level sectors, we encounter coupled blocks of maximum size 15, e.g.,~for the non-planar family corresponding to the non-planar diagram~\ref{fig:vert:2}.
The DEs of 1115 masters are solved numerically with boundary values obtained using the auxiliary mass flow method~\cite{Liu:2017jxz,Liu:2020kpc,Liu:2021wks,Liu:2022mfb} as implemented in \texttt{AMFlow}~\cite{Liu:2022chg}.
Here, it is important to note that since deriving our optimized DE system takes a considerable amount of computation resources, we modified the routines in \texttt{AMFlow} to allow for the direct input of external DEs with symbolic kinematic dependence, thereby avoiding the leading bottleneck of this part of the computation.
%Furthermore, we note that the generation of boundary values performed internally in \texttt{AMFlow} is independent of $m_s$, i.e., only the solution of the main system of DEs depends on the target points in $m_s$-space. 
%Consequently, it is modified to perform this step only once to further eliminate computational overhead. 
The internal generation of the $m_s$-independent boundary values for our system is performed only once to further eliminate computational overhead.
These modifications prove to be vital for us to obtain the high-precision evaluations of all three-loop masters (in the whole $m_s$ domain with 100 significant digits) on our computing facilities. 
The numerical results so-derived have been only checked against the direct LME results but with good agreement. %(in the large-mass limit). %%In fact we could not obtain results for all our 3-loop masters (at one $m_s$ point) with the original un-modified \texttt{AMFlow} on our high-performance workstations in 3 weeks.

\section{Results and discussions}
\label{sec:rest}

Having computed the perturbative results for both $Z_{m}$ and $Z_{\overline{m}}$ determined up to three loops, we then extract $\mathrm{C}_m$ from their ratio $Z_{m}/Z_{\overline{m}}$ perturbatively expanded in $\MSbar$-renormalized $\alpha_s$ followed by taking the $\epsilon \rightarrow 0$ limit.
The perturbative result for $\mathrm{C}_m$ can be parameterized as 
%\begin{equation}\label{eq:Cm_asexp}
$\mathrm{C}_m 
 = 1 + C_m^{(1)}\, \alpha_s/\pi + C_m^{(2)} \, \big(\alpha_s/\pi\big)^2 + C_m^{(3)}  \big(\alpha_s/\pi\big)^3 \, + \mathcal{O}(\alpha_s^4)\,,$
%\end{equation}
where $C_m^{(i)}$ are functions of the scale variables $\bar{\mu}, \mu_s, m_R$, and depend in polynomial form on $\xi$ as well as QCD color factors $C_A=3\,, C_F=4/3$ and $n_l$. %(the dependence on the latter are of polynomial form and not explicitly indicated for simpler notations).  
As explained in section~\ref{sec:RCmass}, we discuss here only the results in Landau gauge $\xi=0$. 
Following from the definition~\eqref{eq:massconversionrelation}, we can derive the following Renormalization-Group (RG) equation regarding the dependence of $\mathrm{C}_m$ on %the $\MSbar$-renormalization scale 
$\bar{\mu}$:%
\begin{equation}\label{eq:Cm_RGE}
\bar{\mu}^2\frac{\mathrm{d} }{\mathrm{d} \bar{\mu}^2} \mathrm{C}_m = 
\bar{\mu}^2\frac{\partial }{\partial \bar{\mu}^2} \mathrm{C}_m + 
\beta\,\alpha_s \frac{\partial }{\partial \alpha_s} \mathrm{C}_m = \gamma_m \, \mathrm{C}_m
\end{equation}%
where $\beta \equiv \bar{\mu}^2\frac{\mathrm{d} \ln(\alpha_s)}{\mathrm{d} \bar{\mu}^2}$ and $\gamma_m \equiv \bar{\mu}^2\frac{\mathrm{d} \ln{\overline{m}_R(\bar{\mu})}}{\mathrm{d} \bar{\mu}^2}$ denote, respectively, the anomalous dimensions of $\alpha_s$ and $\overline{m}_R$.
With the explicit logarithmic dependence on $\bar{\mu}$ of our results for $C_m^{(i)}$ for $i=1,2,3$ we have verified the above RE equation, which also serves as a welcome check of the non-power-suppressed logarithmic dependent part of $\mathrm{C}_m$ on $\mu_s$ in our results.
Below we will simply set $\bar{\mu} = \mu_s$. %namely the renormalization scale of $\alpha_s(\bar{\mu})$ (and $\overline{m}_R(\bar{\mu})$) equals to the subtraction scale $\mu_s$. 
With the above settings, the mass-dimensionless $C_m^{(i)}$, being polynomials in $C_A\,, C_F$ and $n_l$, have their polynomial coefficients reduced to non-trivial functions of just the ratio $m_s \equiv m^2_R/\mu^2_s$.
It is precisely these univariate coefficient functions that we have determined to high precision.
The results for $\mathrm{C}_m$ keeping full dependence on $C_A\,, C_F$, $n_l$, $\xi$ and $\bar{\mu}$ 
up to three loops are provided in the form of fast-evaluating interpolation functions with about 20 significant digits in the supplemental file. %\footnote{The results in the whole range can be obtained from the authors upon request.}
Our one-loop expression of $\mathrm{C}_m$ fully agrees with the result in ref.~\cite{Boyle:2016wis}.
We have checked explicitly that the perturbative conversion factor $\mathrm{C}_m$ is free of the characteristic leading infrared-renormalon behavior~\cite{Bigi:1994em,Beneke:1994sw,Beneke:1998ui,Komijani:2017vep} up to three loops, and thus (m)SMOM mass is a short-distance mass.

\subsection{Numerical results for $\mathrm{C}_m$ in QCD with $3+1$ quarks}
\label{sec:rest_nums}

\begin{figure}
    \centering
    \includegraphics[width=0.92\linewidth]{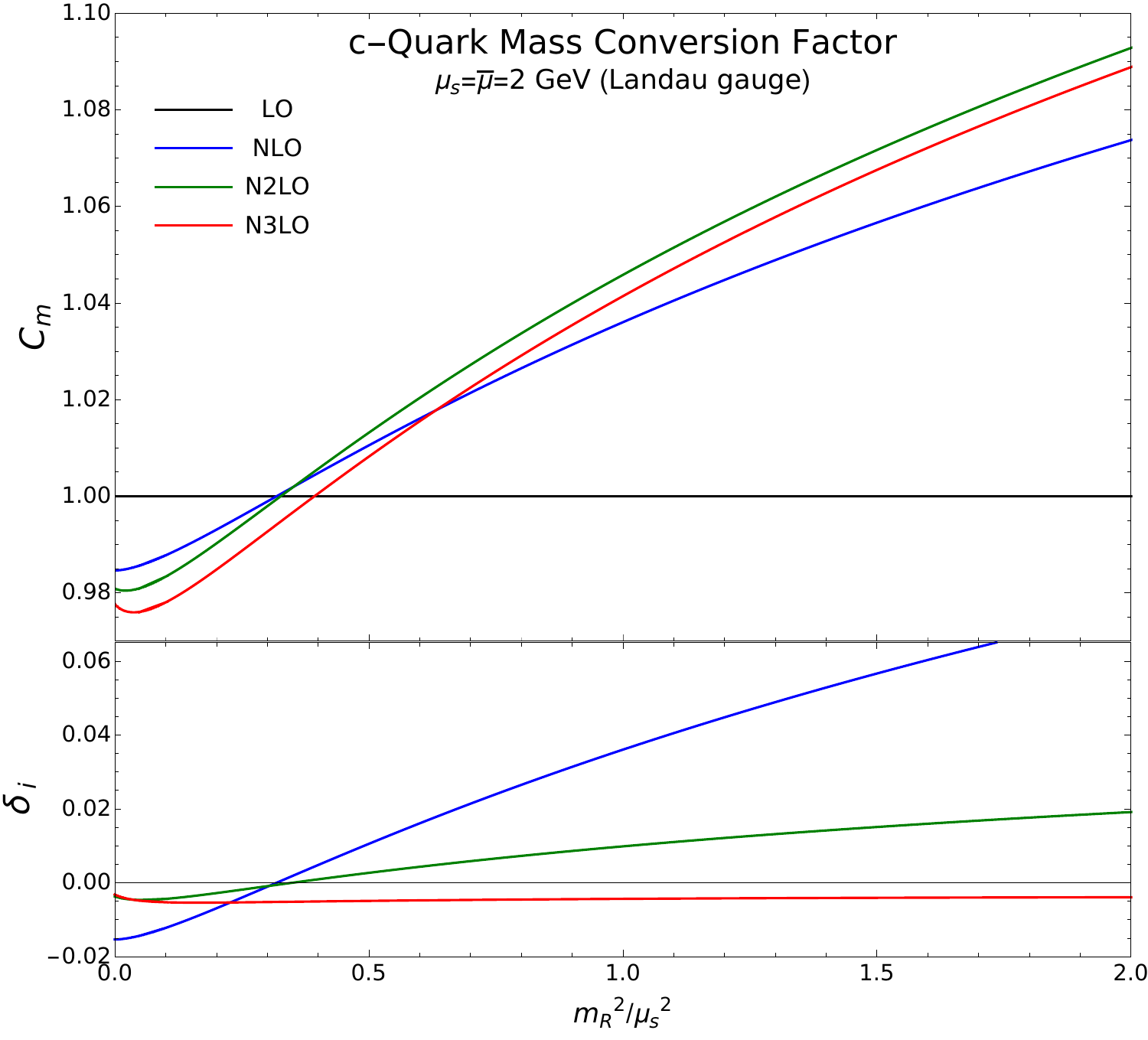}
    \caption{Numerical results for $\mathrm{C}_m$ as functions in $m_s \equiv m^2_R/\mu^2_s$ up to 3-loop order evaluated in QCD with $n_l=3$ at $\bar{\mu}=\mu_s=2$~GeV in Landau gauge}
    \label{fig:Cm_cquark}
\end{figure}

We now focus on the scenario of the $c$-quark mass determination using mSMOM, as recently done in ref.~\cite{DelDebbio:2024hca}.
We consider continuum QCD with only $c$-quark kept massive and $n_l=3$, and work at the matching scale $\mu_s = 2$~GeV, typically chosen for this problem~\cite{FlavourLatticeAveragingGroupFLAG:2024oxs,DelDebbio:2024hca}. 
For definiteness, we set $\alpha_s(\bar{\mu}=\mu_s) = 0.3$ (obtained through 3-loop RG running from input $\alpha_s^{n_f=5} = 0.1179$ at the Z-pole mass to $\bar{\mu}=2$~GeV with the $b$-quark decoupled at its pair-production threshold).
With this setting of parameters, our numerical results for $\mathrm{C}_m$ are plotted in figure~\ref{fig:Cm_cquark} as functions in $m_s \equiv m^2_R/\mu^2_s$ up to three loops.
From the upper panel, we observe that $\mathrm{C}_m$ at three-loop order ranges from about $0.98$ to $1.09$ within $m_s \in (0, 2)$, and all three colored curves intersect with the horizontal tree-level curve within a small region in $m_s$ about $(0.3, 0.4)$.
As becomes clear from the lower panel showing $\delta_i = (\alpha_s/\pi)^i\, C_m^{(i)}$ for $i=1,2,3$ (i.e.~the pure $i$-th loop correction), both $C_m^{(1)}$ and $C_m^{(2)}$ experience a sign flip (at $m_s \sim 0.33$), but not $C_m^{(3)}$.
In fact, the dependence of $C_m^{(3)}$ on $m_s$ is more dramatic than that of $C_m^{(1)}$ and $C_m^{(2)}$, but suppressed in the plot by the factor $(\alpha_s/\pi)^3$.
Furthermore, within $m_s \in (0, 1)$, the typical domain of application for $c$-quark mass determination at $2$ GeV, $\delta_3$ is quite comparable to that of $\delta_2$ in size, albeit differing in sign for $m_s > 0.33$, and hence an explicit determination of these corrections as done here is necessary for providing a more accurate mass conversion from mSMOM to $\MSbar$ and more reliable estimate of the perturbative truncation errors. 
A preliminary application of our $C_m$ to the numbers in ref.~\cite{DelDebbio:2024hca} shows that the central value of the RBC/UKQCD'24 result for $c$-quark $\MSbar$ mass at $2$ GeV could be reduced by about $1\%$ in relative, becoming more consistent with the FLAG averages~\cite{FlavourLatticeAveragingGroupFLAG:2021npn,FlavourLatticeAveragingGroupFLAG:2024oxs}.

It is also interesting to examine the chiral limit $m_s \rightarrow 0$ (the left end of plot~\ref{fig:Cm_cquark}), which precisely corresponds to the results in SMOM. 
Employing a truncated power-logarithm series ansatz, %learned from the solution of the differential equation in quark mass, 
our extrapolated values at the chiral limit agree with the exact SMOM values determined in ref.~\cite{Bednyakov:2020ugu} with more than 85 digits.
This not only serves as a strong check of our computation of $\mathrm{C}_m$ away from the chiral limit, but also provides a lower bound for the numerical precision of our results. 
We conclude by pointing out that there exists a domain, about $m_s \in (0.09, 0.61)$, where the difference of $\mathrm{C}_m$ from $1$ is smaller than the SMOM counterpart up to three loops, potentially making it advantageous in reducing the systematic errors when extracting the $c$-quark $\overline{\mathrm{MS}}$ mass from lattice results.

\subsection{An alternative interpretation of mSMOM conditions in DR}
\label{sec:rest_acrobatic}

The result for $\mathrm{C}_m$ presented in the previous subsection was derived in the standard way of interpreting the renormalization conditions~\eqref{eq:massRCs_f} and~\eqref{eq:massRCs_m} as exact equations in DR, i.e.~to all orders in $\epsilon$, with field, mass and operator renormalization constants Laurent-expanded \textit{necessarily} to sufficiently high orders in $\epsilon$. 
In particular,~\eqref{eq:massRCs_f} and~\eqref{eq:massRCs_m} are solved for 
the parametric coefficients in $Z_m =  1 + \sum_{i=1}^{3} \sum_{j=-i}^{3-i} Z_m^{(i\,j)}\, \alpha_s^i\,\epsilon^j + \mathcal{O}(\alpha_s^4)$ (and similarly $Z_q$) up to three loops in our calculation. 
The higher the loop order in question, the deeper the expansion in both $\alpha_s$ and $\epsilon$.
In view of the subtlety possibly encountered when taking into account the singlet-type diagrams with $\gamma_5$ in closed fermion chains\footnote{These diagrams are irrelevant in SMOM provided one considers only non-singlet axial or pseudoscalar operators.} (see, e.g.~refs.~\cite{Larin:1993tq,Chen:2023lus,Chen:2020ykl}), 
we explore an illuminating possibility that, in the continuum QCD in DR, the original (m)SMOM conditions for all bilinear quark operators, irrespective of involving $\gamma_5$ or not, can also be interpreted merely in a weak sense, namely as equations holding just in the 4-dimensional limit rather than exactly in $d$ dimensions.
To be more specific, for determining $\mathrm{C}_m$ we propose the following variant:
\begin{eqnarray} 
\frac{1}{4 N_c \, p^2} \mathrm{Tr}\big[i\, S^{-1}_R(p, m_R)\, \slashed{p}\big]\Big|_{p^2 = -\mu_s^2;\, \epsilon \rightarrow 0 }\, =\, 1\,, 
\nonumber\\
\frac{1}{4 i N_c} \mathrm{Tr}\big[ \Lambda_{P\,,R}\, \gamma_5\big]\Big|_{p_1^2 = p_2^2 = q^2 = -\mu_s^2;\, \epsilon \rightarrow 0 }\, =\, 1\,, \label{eq:massRCs_4dim}
\end{eqnarray}
to be solved for a smaller number of parametric coefficients defined by $\tilde{Z}_m = 1 + \sum_{i=1}^{3} \sum_{j=-i}^{0} \tilde{Z}_m^{(i\,j)}\, \alpha_s^i\,\epsilon^j + \mathcal{O}(\alpha_s^4)$ and $\tilde{Z}_q = 1 + \sum_{i=1}^{3} \sum_{j=-i}^{0} \tilde{Z}_q^{(i\,j)}\, \alpha_s^i\,\epsilon^j + \mathcal{O}(\alpha_s^4)$.
Note that by construction the $\epsilon$-expansion of each perturbative coefficient is always truncated to $\mathcal{O}(\epsilon^0)$ irrespective of the loop order in question or the leading poles of the bare amplitudes. %The Laurent series expansion of the bare amplitudes shall be done to the order depending on the leading poles of these multiplicative renormalization constants.
Consequently, the higher the loop order in question, the less number of $\tilde{Z}_m^{(i\,j)}$ and $\tilde{Z}_q^{(i\,j)}$ are involved in~\eqref{eq:massRCs_4dim} as compared to those in the usual original interpretation~\eqref{eq:massRCs_f} and~\eqref{eq:massRCs_m}.

What we find is that the weaker variant~\eqref{eq:massRCs_4dim} does work, albeit, leading to different but simpler mSMOM field and mass renormalization constants $\tilde{Z}_q$ and $\tilde{Z}_m$ (i.e.~with different $\epsilon$ coefficients\footnote{Incorporation of the $\epsilon$-power-suppressed terms in the series ansatz for $Z_m$ and $Z_q$ will affect the finite $\mathcal{O}(\epsilon^0)$-terms of the renormalized amplitudes, which can contain contributions from the product of $\epsilon$-poles in the unsubtracted amplitudes and the $\epsilon$-suppressed terms in $Z_m$ and $Z_q$. This will alter the system of algebraic equations extracted for the $\epsilon$-expansion coefficients, subsequently resulting in solutions different from the usual computational prescription.}) from those defined in section~\ref{sec:RCmass}. However, the very same finite $\mathrm{C}_m = \tilde{Z}_m/Z_{\overline{m}}|_{\epsilon \rightarrow 0}$ as presented in the previous subsection still follows in the end.
We anticipate that this point is not limited to the particular renormalization condition or renormalized quantity in question, holding true with broader applicability.  %as long as $\MSbar$ renormalization works.

In addition, we have carried out a further interesting exercise, inspired by refs.~\cite{Liu:2022mfb,Liu:2022chg}.
Instead of inserting the $\epsilon$-expansions of the bare $\Lambda_{P\,,B}$ and $S_B$ into~\eqref{eq:massRCs_4dim} and subsequently extracting the $\epsilon$-free \textit{exact} algebraic equations for $\tilde{Z}_q^{(i\,j)}$ and $\tilde{Z}_m^{(i\,j)}$ (truncated precisely to $\mathcal{O}(\epsilon^0)$), 
we directly substitute the values of $\Lambda_{P\,,B}$, $S_B$ as well as $\tilde{Z}_q$, $\tilde{Z}_m$ at a set of $\epsilon$-points~\cite{Liu:2022mfb,Liu:2022chg}~(in our case $10^{-4} + n \times 10^{-5}$ for $n=1,\cdots, 5$), and then solve~\eqref{eq:massRCs_4dim} for $\tilde{Z}_q^{(i\,j)}$ and $\tilde{Z}_m^{(i\,j)}$ at each chosen $\epsilon$-point.
Note that the symbolic expansion and truncation in $\epsilon$ can no longer be done when solving~\eqref{eq:massRCs_4dim} this way, and the algebraic equations so-extracted for $\tilde{Z}_q^{(i\,j)}$ and $\tilde{Z}_m^{(i\,j)}$ are, in fact, \textit{not} strictly correct in the sense that they are not the same as those algebraic equations derived in the previous treatment.
However, the same numerical result for $\mathrm{C}_m$ can be again restored by extrapolating to the limit $\epsilon \rightarrow 0$, conceptually similar as continuum extrapolation in Lattice calculations.

The observations highlighted above are not merely intriguing but also carry substantial practical significance.
Firstly, $\tilde{Z}_q$ and $\tilde{Z}_m$ in our weaker variant~\eqref{eq:massRCs_4dim} has no $\epsilon$-suppressed terms by construction, and consequently with less number of unknown variables in the fit, the higher numerical precision can be achieved with less computational effort. %\footnote{For our computation here we utilized the fact that $Z_{\overline{m}}$ is known already.}
%this advantage shall become even more pronounced at higher loop orders.
Secondly, when incorporating the singlet-type diagrams in matrix elements of axial-current and pseudoscalar operators in mSMOM, where a single $\gamma_5$ appears on a closed fermion chain, the subtle treatment of $\gamma_5$ in DR shall no longer pose a conceptual issue based on the above investigation.

We conclude this subsection by a brief comment on the renormalization of the scalar operator in mSMOM. 
We have considered both the original mSMOMS condition $\mathrm{Tr}\big[ \Lambda_{S\,,R}\,- 4i/q^2 m_R \Lambda_{P\,,R} \gamma_5 \slashed{q}\big] \big|_{\mathrm{sym}} = 4 N_c\,$, i.e.~eq.~(25) of ref.~\cite{Boyle:2016wis} (where $\Lambda_{S\,,R}$ denotes the renormalized 1PI scalar vertex function and the subscript $\mathrm{sym}$ denotes $p_1^2 = p_2^2 = q^2 = -\mu_s^2$), 
and a simple SMOM-alike condition $\mathrm{Tr}\big[ \Lambda_{S\,,R}\big]\big|_{\mathrm{sym}} = 4\,N_c$ but evaluated at finite non-zero $m_R$.
We have obtained two different $Z_S$ factors, which all differ from $Z_P = Z_m$ in general. 
However, we have verified explicitly that both converge to $Z_P$ in the limit $m_R \rightarrow  0$, as expected based on the restoration of chiral symmetry in massless limit.
We agree with ref.~\cite{Boyle:2016wis,DelDebbio:2024hca} that only in Feynman gauge ($\xi=1$), $Z_S$ determined using eq.~(25) of ref.~\cite{Boyle:2016wis} happens to coincide with $Z_m$ at one loop, but we have checked explicitly that this, unfortunately, does \textit{not} hold beyond that.
On the other hand, since the scalar operator is neither involved in the vector- nor axial-current Ward-Takahashi identities, its renormalization completely decouples from those of other bilinear quark operators. 
Hence we may exploit this freedom to take them equal in the renormalization prescription from the outset. 

\section{Conclusion}
\label{sec:conc}

LQCD calculations employing SMOM and its extension mSMOM -- promising in reducing heavy-quark mass-dependent discretization effects -- are playing an increasingly critical role in precise determination of quark masses, and high-order perturbative results for matching to the commonly used $\overline{\mathrm{MS}}$ masses are among the indispensable ingredients.
We have successfully overcome the formidable challenge of computing three-loop massive off-shell vertex functions and present, for the first time, the result for the quark-mass conversion factor $\mathrm{C}_m$ from mSMOM 
to $\overline{\mathrm{MS}}$ up to three loops in QCD, 
extending the existing result by two additional orders.
Focusing on the $c$-quark mass determination, our results for $\mathrm{C}_m$ exhibit rich behaviors and, notably, there exists a domain in the subtraction scale $\mu_s$ and $m_R$ where $\mathrm{C}_m$ receives consistently less QCD corrections than its SMOM counterpart up to three loops. 
A preliminary application of our $\mathrm{C}_m$ further improves the agreement between the recent RBC/UKQCD'24 determination of the $c$-quark $\MSbar$ mass and its previous FLAG averages.

In addition, we have explored an illuminating possibility that in DR the original (m)SMOM renormalization conditions may also be interpreted merely in a weak sense, namely as equations holding just in the 4-dimensional limit rather than exactly in $d$ dimensions:
they result in different, albeit simpler, DR renormalization constants but still the same finite quark-mass conversion factor.
Moreover, when solving the proposed weaker variant of the (m)SMOM conditions, an exact truncation to the finite order in $\epsilon$ %, i.e.~dropping $\epsilon$-suppressed terms, 
is not necessary. 
We further show that the two renormalization conditions considered for the scalar operator yield factors differing from the pseudoscalar case beyond one loop, but both converge to the latter in the chiral limit. 

There are several interesting follow-ups, which include investigating the QCD$\oplus$QED mixed corrections; 
%derive a RG-improved $\mathrm{C}_m$ by resumming all non-power-suppressed logarithmic dependence on $\mu_s$; %(in the situation where the ratio of $\mu_s/m_R$ is not close to 1) then subtracting the resummed logarithmic part from the fixed-order expressions that contains the full power-suppressed corrections.
supplementing the existing (m)SMOM conditions with an additional RI renormalization condition such that the $\xi$-dependent $\mathrm{C}_m$ becomes useful beyond the Landau gauge; 
extending the setup to study the mass effects of other quarks, for both flavor-(non)singlet bilinear quark operators and the bilinear tensor-current operators, as well as more importantly the four-quark operators relevant for studying meson mixing and hadron decays.

The significance of our work extends beyond its record-setting precision and immediate utility for precision phenomenology. 
The novel and compelling observations uncovered in our analysis may not only help to refine the application of RI/(m)SMOM renormalization, but also be useful in reducing considerably the computational effort especially when incorporating additional quarks with different masses, such as in future applications to $b$-quark mass determinations, and paving the way to incorporate higher-order singlet-type diagrams.

\section*{Acknowledgements}

The work of L.~C. was supported by the Natural Science Foundation of China under contract No.~12205171, No.~12235008, No.~12321005, and by the Taishan Scholar Foundation of Shandong province (tsqn202312052) and 2024HWYQ-005.
The authors gratefully acknowledge the valuable discussions and insights provided by the members of the China Collaboration of Precision Testing and New Physics. 
We thank also Xiao Liu and Xin Guan for helpful discussions regarding certain technical aspects of the computation of loop integrals involved. 
All Feynman diagrams have been produced with the help of \texttt{FeynGame}%~\cite{Harlander:2020cyh,Harlander:2024qbn,Bundgen:2025utt}.
~\cite{Harlander:2020cyh,Bundgen:2025utt}.

\bibliographystyle{utphysM}
\balance
\biboptions{sort&compress}

\bibliography{MS2RImSMOM} 

% ********** Ending **********
\end{document}